\documentclass[10pt,preprintnumbers,amsmath,amssymb,twocolumn]{revtex4}
\usepackage{graphicx}
\usepackage{bm} 

\begin{document}
\title{The Giant Dipole Resonance as a quantitative constraint
on the symmetry energy}
\author{Luca Trippa, Gianluca Col\`o and Enrico Vigezzi\\
Dipartimento di Fisica, Universit\`a degli Studi and INFN, 
Sezione di Milano, via Celoria 16, 20133 Milano, Italy}

\begin{abstract}
The possible constraints on the poorly determined symmetry 
part of the effective nuclear Hamiltonians or effective 
energy functionals, i.e., the so-called symmetry energy $S(\rho)$, 
are very much under debate. In the present work, we show 
that the value of the symmetry energy associated with Skyrme 
functionals, at densities $\rho$ around 0.1 fm$^{-3}$, is 
strongly correlated with the value of the centroid of the Giant Dipole 
Resonance (GDR) in spherical nuclei. Consequently, the 
experimental value of the GDR in, e.g., $^{208}$Pb can be used
as a constraint on the symmetry energy, leading to 23.3 MeV $< 
S(\rho=0.1\ {\rm fm}^{-3}) <$ 24.9 MeV.
\end{abstract}   

\maketitle

\section{Introduction}

The nuclear structure community is currently striving 
to determine a nuclear energy functional as universal and as 
accurate as possibile. Extraction of this functional from a 
more fundamental theory like QCD is of course desirable, and 
there has been progress along this line. At present, however, 
it is still unavoidable to work with functionals which
depend on free parameters which have to be determined by some
fitting procedure.

Existing functionals include those based on a covariant 
formulation~\cite{RMF}
as well as those based on nonrelativistic formulations.
Restricting ourselves to the latter case, we note
that a nuclear energy functional can be defined in a 
general way, without being derived from an underlying 
Hamiltonian in conjunction with a reference state.
However, most of the existing functionals to date 
are derived from an effective Hamiltonian $H_{\rm eff}$
which includes the kinetic energy plus a two-body interaction.
In this case, the total energy is the expectation value of
$H_{\rm eff}$ over the most general Slater determinant
$\vert\Phi\rangle$. Both zero-range interactions like
the one proposed by Skyrme at the end of the 
fifties, and systematically parametrized for Hartree-Fock 
(HF) calculations since the seventies~\cite{Skyrme,SkyrmeOrsay}, 
or finite-range interactions like the Gogny 
force~\cite{Gogny}, lead to 
satisfactory descriptions of many nuclear properties.

Our tool of choice in the present work is the zero-range Skyrme
force, from which one can derive a functional ${\cal E}[\rho]$
which is a function of {\em local} densities only.
For a system that is not symmetric in neutrons and protons, 
the total energy depends both on neutron and proton density:
\begin{equation}
E[\rho] = \int d_3r\ {\cal E}(\rho_n(\vec r),\rho_p(\vec r)).
\end{equation}
For the sake of simplicity, we have not indicated that 
in general the energy
depends not only on the spatial densities, but also on
gradients $\nabla\rho_q$, on the kinetic energy densities
$\tau_q$ and on the spin-orbit densities $J_q$ (where $q$ 
labels $n,p$)~\cite{Chabanat,Bender}. 

In infinite matter, one has a simple expression in terms 
of the spatial densities only. Instead of $\rho_n$ and 
$\rho_p$, one can use the total density
$\rho$ and the {\em local} neutron-proton asymmetry,
\begin{equation}
\delta \equiv {\rho_n-\rho_p \over \rho}.
\end{equation}
This quantity should not be confused with the {\em global} 
asymmetry $(N-Z)/A$. In asymmetric matter, we can 
make a further simplification on ${\cal E}(\rho,\delta)$ by 
making a Taylor expansion in $\delta$ and retaining only 
the quadratic term,
\begin{eqnarray}\label{def_sym}
{\cal E}(\rho,\delta) & \approx & {\cal E}_0(\rho,\delta=0) +
{\cal E}_{\rm sym}(\rho) \delta^2 \nonumber \\
& = & {\cal E}_0(\rho,\delta=0) +
\rho S(\rho) \delta^2.
\end{eqnarray}
The first term on the r.h.s. is the energy density of
symmetric nuclear matter ${\cal E}_{\rm nm}$, while the 
second term defines the main object of the present study, 
namely the {\em symmetry energy} $S(\rho)$. The symmetry 
energy at saturation $S(\rho_0)$ is denoted by different 
symbols in the literature: $J$, $a_\tau$ or $a_4$. We stress 
that Eq.~(\ref{def_sym}) is not really a simplification: the 
coefficient of the term in $\delta^4$ which should follow, 
for the Skyrme parameter sets employed in this work, is 
negligible at densities of the order of $\rho_0$. 
We remind that the pressure of the system can be written
in a uniform system as
\begin{equation}
P = - \left. {\partial E \over \partial V} \right|_A = 
\left. \rho^2 {\partial \over \partial \rho}{{\cal E} \over 
\rho} \right|_A.
\end{equation}
This quantity is evidently related to the density dependence
of the energy functional and of the associated symmetry
part defined above.

The magnitude and the density dependence of the symmetry 
energy $S(\rho)$ are not yet well understood ~\cite{steiner}.
In brief, there exist at present three main research lines 
aimed at constraining the behavior of the symmetry energy,
by using either nuclear structure data, or observables
related to heavy-ion collisions, or evidences from
the study of neutron stars. 

Within the realm of nuclear structure, the symmetry energy
affects of course all properties of nuclei having neutron
excess, including  basic ones like masses and radii. In 
particular, much attention has been focused on radii since 
Typel and Brown~\cite{brown,typelbrown} have noted that 
the neutron skin thickness $\delta R \equiv \langle r^2_n 
\rangle^{1/2} - \langle r^2_p \rangle^{1/2}$ is correlated with 
$P_{\rm nm}(\rho=0.1)$ (see also~\cite{furnstahl}). The 
issue is also investigated in Ref.~\cite{sagawa_r}, where the
correlations between the neutron skin thickness and other
quantities are discussed. The experimental 
accuracy is not sufficient (so far) to limit the
acceptable range of the neutron skin thickness so that this
can constrain a given equation of state; the Parity Radius
Experiment (PREX) at JLAB promises to achieve this
task~\cite{prex}. Another interesting way to fix the value
of the neutron skin thickness, and extract information on
the symmetry energy, is to go through the study of the 
isovector spin-dipole resonance (SDR) sum rule; also in
this case, the experimental difficulties hinder a too
definite conclusion (see Ref.~\cite{sagawa_sdr}).  

In this paper we shall instead concentrate on the correlation between 
the symmetry energy and the energy of the Giant Dipole 
Resonance (GDR). The idea is not new, but in the
present work we develop it based on a fully microscopic
approach, namely within a self-consistent Random Phase Approximation 
(RPA) scheme to calculate the GDR properties. In the 
past~\cite{Krivine}, as well as in recent works~\cite{Danielewicz}, 
the connection with the symmetry energy has been discussed 
starting from a macroscopic, hydrodynamical description
of the GDR (in particular, using the Steinwedel-Jensen 
ansatz which is known to be not fully reliable). Consequently,
we believe that our results are more relevant from
a quantitative point of view. 

If one tries to constrain the symmetry energy by means 
of the study of heavy-ion collisions, or by 
neutron star observables, it is likely that somewhat different
physics is involved. In general, the behavior of the symmetry
energy on a {\em broader} range of densities is involved. 
In heavy-ion collisions maximum densities up to $\sim$4-5 $\rho_0$ 
can be attained~\cite{science}. On the other hand, 
data at lower incident energies are believed to be able
to constrain the nuclear EOS below $\rho_0$~\cite{ono}.  
The study of neutron stars, as well, brings in the physics 
of both low-density and high-density neutron matter (for
a comprehensive review, cf. Ref.~\cite{report}). A recent 
study~\cite{stone} which is close in spirit to ours, 
has examined a large set of Skyrme forces, trying to
determine those which have a satisfactory behavior in 
reproducing the neutron-star observables.
One should remark, however, that there 
are many caveats in the literature against the use of
functionals in a density regime far from that in which
the functionals are fitted and usually employed. In 
particular, we mention here that Monte Carlo calculations 
of neutron matter at low density~\cite{panda} show that
in this regime $E/A$ is about one half of the Fermi energy of
the non-interacting neutron gas, and this behavior is not 
reproduced by any effective mean field functional. 

Consequently, we do not discuss in detail in the present work
the possibility of an overall constraint on the symmetry energy
extracted by different kinds of studies. We briefly discuss 
in the conclusions to what extent our results can be compared 
with a few others in the literature.

\section{The correlation between the GDR and the symmetry energy}

Our starting point will be the hydrodynamical model of giant
resonances, proposed by Lipparini and Stringari~\cite{ls}. They assume
an energy functional which is simplified yet sufficiently realistic,
solve the macroscopic equations for the densities and currents, 
and extract expressions for the moments $m_1$ and $m_{-1}$ 
associated with an external operator $F$ 
($m_k\equiv\int dE S(E) E^k$ where $S$ is the strength 
function associated with $F$). The expression for $m_1$ 
is proportional to $(1+\kappa)$, where $\kappa$ is the 
well-known ``enhancement factor'' which in the case 
of Skyrme forces is associated with their velocity
dependence. The expression for $m_{-1}$, in the case of an
isovector external operator, includes integrals 
involving ${\cal E}_{\rm sym}$ and $F$. They can be 
evaluated in a simple way if one assumes the validity of 
the leptodermous expansion. We write the volume and 
surface coefficients of the expansion of ${\cal E}_{\rm sym}$ 
as $b_{\rm vol}$ and $b_{\rm surf}$, respectively.
By specializing $F$ to the isovector dipole case, the 
following expression is obtained (for details, cf. 
Ref.~\cite{ls})
\begin{equation}\label{Els}
E_{-1}\equiv\sqrt{\frac{m_1}{m_{-1}}}
=\sqrt{\frac{3\hbar^2}{m\langle 
r^2\rangle}\frac{b_{\rm vol}}{\left[1+\frac{5}{3}\frac{b_{\rm surf}}
{b_{\rm vol}}A^{-\frac{1}{3}} \right]}(1+\kappa)}.
\end{equation}
This equation yields values of the centroid energy which 
are in rather good agreement with those of microscopic 
RPA calculations. It turned out to be useful in a previous 
study~\cite{plb95}, in order to constrain directly the 
parameters of the isovector part of the Skyrme interaction.
Here we shall use it as a guideline, and try instead to find 
a quantitative connection between the energy of the GDR
and the symmetry energy.

\begin{figure}[hbt]
\includegraphics[width=6cm,angle=-90]{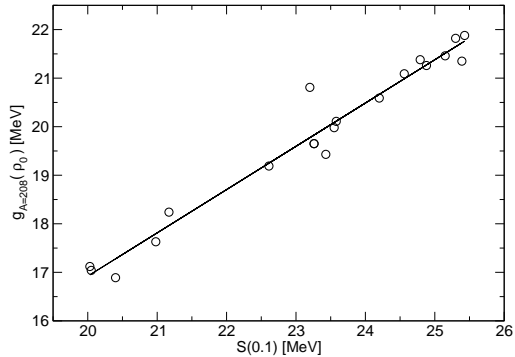}
\caption{Correlation between the quantity $g_{A=208}(\rho_0)$ 
and the symmetry energy $S(\rho)$ for the Skyrme forces listed 
in Table~\ref{table1}, at $\rho= 0.1$ fm$^{-3}$. The value of the 
correlation coefficient is $r = $ 0.981.}
\label{fig_added}
\end{figure}

The ratio $\frac{b_{\rm surf}}{b_{\rm vol}}$ can be evaluated 
through the calculation of a semi-infinite nuclear slab. 
This has been done, e.g., in Ref.~\cite{krivine} (cf. 
their Sec. 3.2.3). We do not discuss here the approximations made
in the derivation, but we use the fact that the mentioned ratio
can be written in terms of the symmetry energy and its derivatives.
If we insert this result into Eq.~(\ref{Els}) we obtain
\begin{equation}\label{Els2}
E_{-1}=\sqrt{\frac{6\hbar^2}{m\langle r^2\rangle}
g_A(\rho_0) (1+\kappa)},
\end{equation}
where
\begin{equation}
g_A(\rho)=
\frac{S(\rho)}{1+\frac{5}
{S(\rho)}[\rho\frac{dS}{d\rho}-\frac{\rho^2}{4}
\frac{d^2S}{d\rho^2}]A^{-\frac{1}{3}}}.
\end{equation}
For a given heavy nucleus we can safely consider the first 
of the three factors under the square root as a constant, 
since different Skyrme forces do not vary widely in their 
predictions for $\langle r^2\rangle$. We have evaluated 
the term $g_A(\rho)$ at $\rho=\rho_0$ for $A$=40, 124, 208, and 
for a number of Skyrme forces. We have found that it is 
strongly correlated with the value of $S(\rho)$ in the 
range $\rho = 0.08 - 0.12$ fm$^{-3}$. The specific
case $A$=208 is displayed in Fig.~\ref{fig_added} 
in the case $\rho=0.1$ fm$^{-3}$, for which the  
correlation coefficient is maximum. 

Although we have not been able to deduce this 
correlation in an analytic way from the expression 
of the Skyrme functional, this result, together 
with Eq.~(\ref{Els2}), motivates us to look for a direct 
correlation between the centroid of the GDR and the 
symmetry energy, through the quantity
\begin{equation}\label{frho}
f(\rho) \equiv \sqrt{S(\rho)(1+\kappa)},
\end{equation}
for $\rho\sim$ 0.1 fm$^{-3}$. In the next Section, 
we discuss this correlation for $\rho$ = 0.1 
fm$^{-3}$ in the case of $^{208}$Pb. We also discuss
in some detail the choice of the Skyrme forces that we have 
employed. 

\vspace{5mm}

\section{Results}

We have obtained results for the GDR in $^{208}$Pb by using 
a series of microscopic Hartree-Fock (HF) plus Random Phase 
Approximation (RPA) calculations. Skyrme-RPA theory is well known 
since many years, especially in its matrix formulation.
Recently, we have developed a scheme which is fully 
self-consistent and is discussed in~\cite{comex2}. There 
is no approximation in the residual interaction, in that all 
its terms are taken into account (including the two-body 
spin-orbit and Coulomb interactions). The occupied states are 
determined by solving the HF equations in a radial mesh extending up to 
24 fm. The continuum is discretized since the positive 
energy states are obtained using box boundary conditions. 
Particle-hole (p-h) configurations which constitute the 
basis for the RPA matrix equations are included up to 
typically 60 MeV so that the value of $m_1$ is
at least 98\% of the well-known value obtained from the 
double commutator. In a few cases, due to the instability 
of RPA, we had to resort to Tamm-Dancoff Approximation (TDA) 
calculations but this does not affect
significantly our results. 

The calculations have been performed for a set of 20 Skyrme 
interactions. It is well known that more than 100 Skyrme 
parametrizations have been proposed in the literature, 
including some which have been employed only in limited 
and specific cases. It is hard to define in a clear-cut 
way a ``standard'' subset to be analyzed; however, we 
have decided, in the present context, not to consider 
parametrizations which (a) have an associated $K_\infty$ 
outside of the range 210-270 MeV, in keeping with the conclusions 
reached in Ref.~\cite{gmr} by studying the giant monopole resonance
(GMR) in $^{208}$Pb, (b) reproduce the experimental value of the
GDR in $^{208}$Pb (13.46 MeV~\cite{adndt}) within $\pm$ 2 MeV. 

We have determined our set including forces proposed by 
different groups, and at different times. In this sense
the set can be considered representative enough. In the cases in 
which several forces have been proposed in the same reference, 
we have not included more than two forces, in order to avoid a
too strong bias. The
forces are listed in Table~\ref{table1}; we also provide the 
reference from which the parameter sets have been taken, the 
value for $E_{-1}(RPA) \equiv \sqrt{m_1\over m_{-1}}$ 
obtained from our RPA calculation, and the values of the
quantities  $f(0.1)$, $S(0.1)$ and $\kappa$ (cf. 
Eq.~(\ref{frho})). 

\begin{table}[hbt]
\begin{center}
\begin{tabular}{lccccc}
\hline
& Ref. & $E_{-1}$ & $f(0.1)$      & $S(0.1)$ & $\kappa$   \\
\hline
&      & [MeV]    & [MeV$^{1/2}$] & [MeV]  &              \\
\hline
SkA     & \cite{SkA}      & 15.14  & 6.27  & 23.43 & 0.68 \\
SkM     & \cite{Krivine}  & 13.94  & 5.65  & 23.26 & 0.37 \\
SGI     & \cite{sagawa}   & 14.16  & 5.62  & 20.40 & 0.55 \\
SGII    & \cite{sagawa}   & 13.56  & 5.34  & 20.98 & 0.36 \\
SkM*    & \cite{skmstar}  & 13.89  & 5.64  & 23.26 & 0.37 \\
RATP    & \cite{ratp}     & 15.17  & 6.06  & 23.55 & 0.56 \\
SkT4    & \cite{tondeur}  & 11.47  & 4.86  & 23.58 & 0.00 \\
SkT6    & \cite{tondeur}  & 12.17  & 4.92  & 24.20 & 0.00 \\
Rs      & \cite{pgr}      & 12.76  & 5.20  & 20.05 & 0.35 \\
Gs      & \cite{pgr}      & 12.62  & 5.20  & 20.03 & 0.35 \\
SkI2    & \cite{SkI}      & 12.29  & 5.00  & 21.17 & 0.18 \\
SLy230a & \cite{Chabanat0}& 12.49  & 5.04  & 25.43 & $\sim$ 0 \\
SLy4    & \cite{Chabanat} & 13.40  & 5.45  & 25.15 & 0.18 \\
SLy5    & \cite{Chabanat} & 13.28  & 5.42  & 24.88 & 0.18 \\
SkO'    & \cite{SkOp}     & 13.85  & 5.00  & 22.61 & 0.11 \\
MSk7    & \cite{msk7}     & 12.10  & 4.86  & 24.56 & -0.04\\
v110    & \cite{vv}       & 12.13  & 4.80  & 24.79 & -0.07\\
v075    & \cite{vv}       & 13.97  & 5.62  & 25.30 & 0.25 \\
SK255   & \cite{shlomo}   & 13.98  & 5.94  & 25.39 & 0.39 \\
LNS     & \cite{lombard}  & 13.95  & 5.43  & 23.20 & 0.27 \\
\hline                                          
\end{tabular}                                   
\caption{For the Skyrme parameter sets considered in this 
work, we provide the values of $E_{-1}(RPA)$, $f(0.1)$, 
$S(0.1)$ and $\kappa$. All these quantities are defined in 
the text.}
\label{table1}                           
\end{center}                                    
\end{table}  

\begin{figure}[hbt]
\includegraphics[width=6cm,angle=-90]{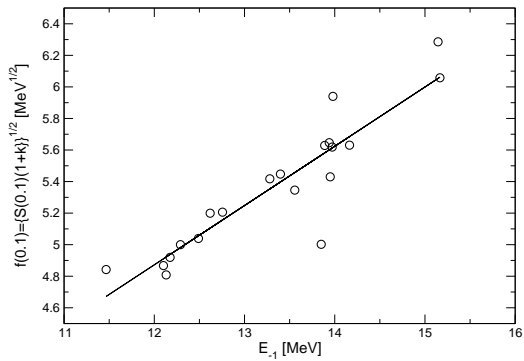}
\caption{Correlation between the energy of the GDR and 
the quantity $f(0.1)$. The definition of these quantities, 
and the related discussion, can be found in the text.}
\label{fig1}
\end{figure}

We find a strong linear correlation between the values of 
$E_{-1}(RPA)$ and $f(0.1)$, which are shown in 
Fig.~\ref{fig1} together with the interpolating straight 
line $f(0.1) =  a + b E_{-1}(RPA)$. The value of the 
correlation coefficient is $r = $ 0.909. Before discussing 
the extraction of the value of the symmetry energy, we should 
stress that we have not been able to correlate 
the GDR simply with $S(\rho_0)$; this
may be possible (cf., e.g., Ref.~\cite{pgnpa}) 
at the price of restricting oneself 
to a small set of Skyrme forces. 

We can now make avail of the experimental values of the GDR centroid 
$E_{-1}(exp)$ and of the enhancement factor $\kappa$ to deduce the 
best value of the symmetry energy. While the value of 
$E_{-1}(exp)$ in $^{208}$Pb has been rather well determined
from photoabsorbtion measurements, $E_{-1}(exp)=$ 
13.46 MeV~\cite{adndt}, the value of $\kappa$
is less precise. Since 
\begin{equation}\label{csk}
\int_0^\infty \sigma(E)dE = 60\left( {NZ \over A} \right)
\cdot(1+\kappa)\ {\rm MeV\cdot mb},
\end{equation}
$\kappa$ would be determined if the integrated photoabsorbtion 
cross section $\sigma(E)$ had been mesured up
to large energies (essentially, up to the pion production threshold).
For obvious experimental difficulties, the photoabsorbtion cross
section has been measured up to much lower energies  
(26.4 MeV in the case at hand, namely $^{208}$Pb~\cite{ves}). 
The uncertainty on $\kappa$ has been already estimated 
in~\cite{ls}, where it has been stated that $\kappa$ should lie, 
approximately, between 0.2 and 0.3. We have checked this in some
detail in the following way. From the experimental cross section
measured up to the maximum energy of 26.4 MeV, that is, from 
Eq. (\ref{csk}) replacing the upper limit in the integral 
with 26.4 MeV, we have deduced an ``effective'' 
$\kappa$ that we can call $\kappa_{26.4}$. The Skyrme forces 
which have an associated $\kappa_{26.4}$ within the experimental 
limit are those which have $\kappa$ between 
$\kappa_{\rm min}$=0.18 and $\kappa_{\rm max}$=0.26. In the
following, we have used these values. Our best value for
$\kappa$ is of course an average between $\kappa_{\rm min}$ 
and $\kappa_{\rm max}$. 

At this point, the best value for $S(0.1)$ is found as
\begin{equation}
S(0.1)=\frac{(a+bE_{-1}^{exp})^2}{(1+\kappa)},
\end{equation}
where $a$, $b$ come from the fit. The error is obtained from
\begin{equation}
\sigma_{\sqrt{S(0.1)(1+\kappa)}}=\sigma_b
\sqrt{\sigma_{E_{-1}}^2+(\bar{E}_{-1}-E_{-1}^{exp})^2},
\end{equation}
where $\sigma_b$ comes from the fit and the variance 
$\sigma_{E_{-1}}^2$ is calculated with respect to the 
interpolating straight line. Having determined the 
$\pm$1$\sigma$ interval around the mean value for the 
quantity $\sqrt{S(0.1)(1+\kappa)}$, we obtain  
\begin{equation}
\frac{5.419}{\sqrt{1+\kappa_{\rm max}}}< \sqrt{S(0.1)} 
< \frac{5.422}{\sqrt{1+\kappa_{\rm min}}}.
\end{equation}
This can be considered one of the main results of the
present investigation. By introducing the values of 
$\kappa_{\rm min}$ and $\kappa_{\rm max}$ discussed
above, one obtains a further (more direct) constrain,  
that is, 
\begin{equation}\label{finalcon}
23.3\;{\rm MeV} < S(0.1) <24.9\;{\rm MeV}.
\end{equation} 

\section{Conclusions}

Using a representative set of Skyrme effective functionals 
we have found a clear correlation between the energy 
of the GDR in $^{208}$Pb and a simple function of 
the symmetry energy at density $\rho \sim 0.1$ 
fm$^{-3}$ and of the enhancement factor $\kappa$ 
associated with the velocity dependence (and with 
the effective mass) of the various functionals.
Using the well established experimental value of the GDR
we have extracted a range of acceptable values of 
for $S(0.1)$ (cf. Eq.~(\ref{finalcon})). 

It would be important to test whether other 
classes of effective functionals lead to a similar result. 
More generally, it will be essential to study the 
interplay between constraints coming from the different 
kinds of works mentioned in the Introduction,
which deal with different energy and density regimes.
In fact, a better knowledge of the symmetry
part of the nuclear effective functionals, 
and in particular of its density dependence, 
would be highly instrumental for the study of 
systems ranging from exotic nuclei to pure neutron matter.

\end{document}